\documentclass[prl,superscriptaddress,twocolumn,showpacs,preprintnumbers]{revtex4}

\usepackage{graphicx}
\usepackage{dcolumn}
\usepackage{bm}

\begin{document}


\title{Nonlinear $\sigma$ Model Method 
for the $J_1$-$J_2$ Heisenberg Model: \\
Disordered Ground State with Plaquette Symmetry} 

\author{Ken'ichi Takano}
\affiliation{Toyota Technological Institute, 
Tenpaku-ku, Nagoya 468-8511, Japan}
\author{Yoshiya Kito}
\author{Yoshiaki \=Ono}
\altaffiliation[Present address: ]
{Department of Physics, Niigata University,
Ikarashi, Niigata 950-2181, Japan.}
\affiliation{Department of Physics, Nagoya University, 
Chikusa-ku, Nagoya 464-8602, Japan}
\author{Kazuhiro Sano}
\affiliation{Department of Physics Engineering, 
Mie University, Tsu, Mie 514-8507, Japan}

\date{24 June 2003}

\begin{abstract}

      A novel nonlinear $\sigma$ model method is 
proposed for the two-dimensional $J_1$-$J_2$ model, 
which is extended to include plaquette-type distortion. 
      The nonlinear $\sigma$ model is properly derived 
without spoiling the original spin degrees of freedom. 
      The method shows that a single disordered phase 
continuously extends from a frustrated uniform regime 
to an unfrustrated distorted regime. 
      By the continuity and Oshikawa's commensurability 
condition, the disordered ground states for the uniform 
$J_1$-$J_2$ model are 
plaquette states with fourfold degeneracy. 

\end{abstract}
\pacs{75.10.Jm, 75.30.Et, 75.30.Kz}

\maketitle


      The two-dimensional (2D) $J_1$-$J_2$ model is 
a frustrated Heisenberg model with nearest neighbor 
(NN) and next nearest neighbor (NNN) antiferromagnetic 
exchange interactions on a square lattice. 
      The model 
with spin magnitude $S=\frac{1}{2}$ is realized 
in mother materials of cuprate superconductors, 
La$_2$CuO$_4$, YBa$_2$CuO$_6$ and 
Sr$_2$CuO$_2$Cl$_2$ as small-$J_2$ 
systems~\cite{Manousakis,Kastner}. 
      Recently found materials, 
Li$_2$VOSiO$_4$ and Li$_2$VOGeO$_4$, 
are also described by the model in the case of 
$J_2/J_1$$\sim$1~\cite{VO_ex,VO_th}. 
	    A particular interest for the $J_1$-$J_2$ model 
is in a gapful disordered state, which may be 
formed by frustration under strong quantum 
fluctuations~\cite{Anderson}. 
      The subject has been theoretically investigated 
by various methods~\cite{Manousakis}: e.~g. 
spin wave theories~\cite{Chandra,Mila,Nishimori}, 
nonlinear $\sigma$ model (NLSM) 
methods~\cite{Chakravarty,Einarsson}, numerical 
diagonalizations~\cite{Dagotto,Sano,Schultz,Capriotti0}, 
quantum Monte Carlo (QMC) 
simulations~\cite{Sorella,Capriotti1,Jongh}, 
series expansions~\cite{Zhitomirsky,Singh,Kotov}, 
and variational methods~\cite{Capriotti2}. 

      For a system only with the NN interactions 
($J_2$=0), the ground state is  believed to have an 
antiferromagnetic (AF) order. 
      The NNN exchange interactions are expected to 
induce strong frustration to break the AF order and 
to form a disordered ground state around 
$J_2/J_1 = 0.5$. 
      A current leading QMC 
calculation~\cite{Sorella,Capriotti1} 
supports the disordered phase with spin-gap for 
$J_2/J_1$$\gtrsim$0.4. 
      Accepting this result, the issue is the character 
of the ground state in the disordered phase. 
      Candidates examined in recent several years are 
a uniform resonating-valence-bond (RVB) 
state~\cite{Capriotti2}, 
a plaquette state~\cite{Zhitomirsky,Capriotti1}, 
a dimer state~\cite{Singh,Kotov,Capriotti2}, 
and a state both with dimer and plaquette 
structures~\cite{Jongh}; 
      their degeneracies are 1, 4, 4 and 8, respectively. 
      Although Oshikawa's commensurability 
condition~\cite{Oshikawa} is useful to 
restrict possibilities, 
it does not completely select one;  
e.~g. it requires that 
a uniform RVB ground state with spin-gap 
is accompanied with gapless singlet excitations. 
      The character of the ground state is still under debate. 

      A disordered state is formed also by 
distortion in the exchange constants, 
even if there is no frustration ($J_2 = 0$). 
      For a plaquette-type distortion, a disordered state 
interpreted as a 2D array of plaquette-singlets 
is formed~\cite{Koga_Voigt}. 
      Here it is a question whether the disordered state 
by frustration is essentially the same as that by 
plaquette-type distortion. 
      If it is the same, a disordered phase continuously 
extends from a regime of strong frustration and weak 
distortion to a regime of weak frustration and strong 
distortion in a parameter space. 
     However, if not, there exists a phase boundary 
between them; then the ground state of the uniform 
$J_1$-$J_2$ model is not plaquette-like. 
      Hereafter we consider the $J_1$-$J_2$ model 
which is extended to include a plaquette-type distortion. 

      Among various methods to analyze spin systems, 
an NLSM method is effective to clarify their characters. 
      The first successful example appeared in 
one dimension. 
      A uniform spin chain with NN interactions is 
mapped onto an NLSM with an appropriate topological 
term~\cite{Haldane1}. 
      Inhomogeneous spin chains 
with periodicity are 
treated by refined and extended NLSM 
methods~\cite{Affleck1,Takano1}. 
      For 2D systems, an NLSM without topological 
term is derived for $J_2 = 0$~\cite{nontopological}. 
      For $J_2 \ne 0$, 
Chakravarty {\it et al.} \cite{Chakravarty} 
analyzed 2D NLSM which 
represents the uniform $J_1$-$J_2$ model. 
      By applying a renormalization group (RG) method to 
the NLSM, they constructed a standard theory for 
the quantum phase transition. 

      Despite the success, there remains 
ambiguity in the correspondence of a derived NLSM to 
the $J_1$-$J_2$ model. 
      If one use a naive mapping in literature, 
a single spin variable is replaced by the sum of 
two new variables representing a slowly varying 
AF motion and a rapid fluctuation. 
      This is not justified because the number of 
independent variables is abruptly increased. 
      Although the mapping may phenomenologically 
produce the correct NLSM, there is no way to confirm 
the correctness within the NLSM method itself. 
      Further the increase of the degrees of freedom 
leaves ambiguity for the choice of the cutoff. 
      In one dimension, the problem of the degrees of 
freedom has been overcome in generalized 
formulations~\cite{Affleck1,Takano1}. 
      However such a reasonable theory in two dimensions 
has not been proposed. 
      To construct a qualified 2D NLSM method is 
a purpose of this Letter. 
      Using the NLSM method, to determine the character 
of the ground state for the 2D $J_1$-$J_2$ model
is the final purpose. 

\begin{figure}[btp]
\begin{center}\leavevmode
\includegraphics[width=0.7\linewidth]{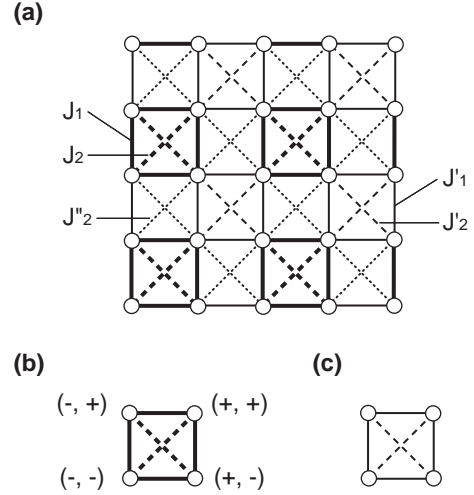}
\end{center}
\vskip -0.3cm
\caption{(a) Lattice of the $J_1$-$J_2$ model. 
      Lattice sites are denoted by small circles and exchange 
constants are by various lines between circles. 
      (b) A plaquette consisting of four sites connected by 
$J_1$ and $J_2$; this is a block which is a unit 
in the NLSM formulation. 
      A variable ${\bf n}_{j}$ in the $p$th block is relabeled 
as ${\bf n}^{\mu\nu}(p)$, where $\mu$ and $\nu$ take 
$+$ or $-$. 
      The value of ($\mu$, $\nu$) at each site is shown. 
      (c) Another kind of plaquette consisting of four sites 
connected by $J'_1$ and $J'_2$. 
} 
\label{plaquette}
\end{figure}

      The $J_1$-$J_2$ model with plaquette-type 
distortion is represented by the Hamiltonian: 
\begin{equation}
H = \sum_{\langle i, j \rangle} 
J_{1;ij} \ {\bf S}_i \cdot {\bf S}_j 
+ \sum_{\langle\!\langle i, k \rangle\!\rangle} 
J_{2;ik} \ {\bf S}_i \cdot {\bf S}_k , 
\label{Hamiltonian}
\end{equation}
where ${\bf S}_i$ is the spin of magnitude $S$ at site $i$.  
      The first and the second summations are taken over
NN and NNN pairs, respectively, in a square lattice. 
      $J_{1;ij}$ takes $J_1$ or $J'_1$,  
and $J_{2;ik}$ does $J_2$, $J'_2$ or $J''_2$ 
as shown in Fig.~\ref{plaquette}(a). 
      The system is reduced to the uniform $J_1$-$J_2$ 
model 
when $J_1$ = $J'_1$ and $J_2$ = $J'_2$ = $J''_2$. 
      In the limit of $J'_1$ = $J'_2$ = $J''_2$ = 0, the 
lattice is an assembly of isolated plaquettes 
each of which consists of four spins connected by 
$J_1$ and $J_2$ (Fig.~\ref{plaquette}(b)). 
      Also, in the limit of $J_1$ = $J_2$ = $J''_2$ = 0, 
the lattice is an assembly of another kind of 
isolated plaquettes; each consists of four spins 
connected by $J'_1$ and $J'_2$ (Fig.~\ref{plaquette}(c)). 
     Hamiltonian (\ref{Hamiltonian}) is 
invariant under the simultaneous exchanges of 
$J_1$ and $J'_1$, and of $J_2$ and $J'_2$. 
      The symmetric case of 
$J_1$ = $J'_1$ and $J_2$ = $J'_2$ 
includes the uniform $J_1$-$J_2$ model. 

      We consider the quantum Hamiltonian 
(\ref{Hamiltonian}) in the classical N\'eel ordered 
region. 
      The expectation value of ${\bf S}_j$ for a spin 
coherent state at imaginary time $\tau$ is given as 
\begin{equation}
\langle {\bf S}_j \rangle = (-1)^j S {\bf n}_j(\tau) 
\ \ {\rm with} \ \ {\bf n}_j^2 = 1 , 
\label{coherent}
\end{equation}
where $(-1)^j$ is a symbol 
taking $+$ or $-$ depending on the sublattice which 
the $j$th site belongs to. 
      The partition function is then written in a path 
integral formula as 
\begin{equation}
           Z =  \! \int \!\! D[{\bf n}_j(\tau)] \, 
\prod_j \delta({\bf n}^2_j(\tau) - 1) \, e^{-A} . 
\label{partition}
\end{equation}
       The action $A$ at temperature $1/\beta$ is given by 
\begin{equation}
      A = i S \sum_{j} (-1)^j w[{\bf n}_j] 
              + \int_0^{\beta}  \! d\tau H(\tau) . 
\label{action_n} 
\end{equation}
      The first term is the Berry phase term 
with the solid angle $w[{\bf n}_j]$ which the unit 
vector ${\bf n}_j(\tau)$ forms in period $\beta$. 
     $H(\tau)$ in the second term is given by 
\begin{eqnarray}
H(\tau) &=& \frac{1}{2} S^2 
\sum_{\langle i, j \rangle} 
J_{1;ij} [{\bf n}_i(\tau) - {\bf n}_j(\tau)]^2
\nonumber \\
&-& \frac{1}{2} S^2 
\sum_{\langle\!\langle i, k \rangle\!\rangle} 
J_{2;ik} [{\bf n}_i(\tau) - {\bf n}_k(\tau)]^2 , 
\label{Hamiltonian_n}
\end{eqnarray}
where the constraint ${\bf n}^2_j(\tau) = 1$ 
in the $\delta$-function of Eq.~(\ref{partition}) has been 
used. 
      Hereafter we do not explicitly write the $\tau$ 
dependence of ${\bf n}_j(\tau)$. 

      We adopt a plaquette of Fig.~\ref{plaquette}(b) as a unit 
of transformation, and call it a {\it block}\,; 
      we would choose another kind of plaquette in 
Fig.~\ref{plaquette}(c) as a block. 
      We relabel four variables, ${\bf n}_{j}$'s, in the $p$th 
block as ${\bf n}^{++}(p)$, ${\bf n}^{+-}(p)$, ${\bf n}^{-+}(p)$ 
and ${\bf n}^{--}(p)$, as shown in Fig. \ref{plaquette}(b). 
      By analogy with the one-dimensional 
case~\cite{Takano1}, we transform them as 
\begin{eqnarray}
      {\bf n}^{\mu \nu}(p) = {\bf m}(p)  
+ a[\mu \nu {\bf L}_{0}(p) 
 + \mu {\bf L}_1(p)  + \nu {\bf L}_2(p)] . 
\label{transform}
\end{eqnarray}
      Here ${\bf L}_0(p)$, ${\bf L}_1(p)$ and ${\bf L}_2(p)$ 
describe small fluctuations around ${\bf m}(p)$. 
      According to the variable transformation, 
four original constraints, 
$[{\bf n}^{\mu \nu}(p)]^2$ = 1 ($\mu, \nu$ = $\pm$), 
are changed to four new 
constraints, ${\bf m}^2(p)$=1 and 
${\bf m}(p)$$\cdot$${\bf L}_q(p)$ = 0 
($q$ = 0, 1, 2). 
      Thus we obtained a new set of variables, the number 
of which is the same as that of the original variables. 
      This plaquette-based transformation is inevitable 
to keep the original degrees of freedom 
even in the uniform $J_1$-$J_2$ model. 

      In the continuum limit, the first term 
of the action (\ref{action_n}) is written as  
$ iS \sum_p \sum_{\mu,\nu} 
                   \mu\nu w[{\bf n}^{\mu\nu}(p)] $
      = $ i(S/a) \int \!\! d\tau d^2{\bf r} \, 
                 {\bf L}_{0} 
\! \cdot \! ({\bf m} \! \times \! \partial_{\tau} {\bf m})$ 
with lattice spacing $a$. 
      For the second term of Eq.~(\ref{action_n}), 
we substitute Eq.~(\ref{transform}) into 
Eq.~(\ref{Hamiltonian_n}) and take the continuum limit. 
      Thus, to the leading order of derivatives and 
fluctuations, we have the field-theoretic action 
\begin{eqnarray}
A &=& S^2\int d{\tau}d^2{\bf r} \Bigl\{
   \frac{i}{Sa} \, {\bf L}_{0} \cdot 
({\bf m} \! \times \! \partial_{\tau} {\bf m}) \Bigr. 
\nonumber \\ 
 &+& J'_0 [(\partial_x{\bf m})^2 
+ (\partial_y{\bf m})^2 
-2\partial_x{\bf m}\cdot{\bf L}_1 
-2\partial_y{\bf m}\cdot{\bf L}_2] 
\nonumber \\ 
 &+& \Bigl. 2(J_1+J'_1) {\bf L}_{0}^2 
    + (J_0+J'_0) ({\bf L}_1^2 + {\bf L}_2^2) \Bigr\} 
\label{action_mL} 
\end{eqnarray}
with 
$J_0 \equiv J_1-J_2-J''_2$ and 
$J'_0 \equiv J'_1-J'_2-J''_2$. 
      This action includes all the low-energy 
excitations surviving the continuum approximation, 
since the original degrees of freedom are not spoiled 
in the variable transformation (\ref{transform}). 
      In Eq.~(\ref{action_mL}), ${\bf L}_0$, ${\bf L}_1$ 
and ${\bf L}_2$ are massive fields \cite{classical}, 
so that they are irrelevant to a symmetry change of 
the ground state. 

      Now we integrate out the partition function for 
the action (\ref{action_mL}) with respect to 
massive fields ${\bf L}_0$, ${\bf L}_1$ and ${\bf L}_2$. 
      The resultant partition function 
contains the NLSM action: 
\begin{eqnarray}
A_{\rm eff} &=& \int \! d\tau d^2{\bf r} 
\biggl\{ \frac{1}{8a^2(J_1+J'_1)} (\partial_{\tau}{\bf m})^2 
\nonumber \\ 
&+& \ S^2 
\left( \frac{1}{J_0}+\frac{1}{J'_0} \right)^{\!\! -1} 
[(\partial_x{\bf m})^2 + (\partial_y{\bf m})^2] 
\biggl\} . 
\label{action_eff} 
\end{eqnarray}
      There appears no topological term  even if the NNN 
interactions exist. 
      The bare spin wave velocity is read as $v$ = 
$2\sqrt{2}Sa(J_1+J'_1)^{1/2}(1/J_0 + 1/J'_0)^{-1/2}$. 
      Action $A_{\rm eff}$ keeps the original invariance 
against the simultaneous exchanges of 
$J_1$ and $J'_1$, and of $J_2$ and $J'_2$, 
meaning that the same action is obtained if 
we use a plaquette in Fig.~\ref{plaquette}(c), 
instead of Fig.~\ref{plaquette}(b), as a block. 
      This result reflects that the variable 
transformation (\ref{transform}) 
does not restrict the spin motion to form a singlet 
on the plaquette of Fig.~\ref{plaquette}(b). 

      We apply the RG analysis 
by Chakravarty {\it et al.} \cite{Chakravarty} 
to the present NLSM. 
      We first introduce rescaled dimensionless coordinates, 
$x_0 = \Lambda v \tau$, $x_1 = \Lambda x$ and 
$x_2 = \Lambda y$, 
with a momentum cutoff $\Lambda$ of order $a^{-1}$. 
      The NLSM action (\ref{action_eff}) 
is then rewritten as 
\begin{eqnarray}
A_{\rm eff} = \frac{1}{2g_0} \int \! d^3 x 
\left( \frac{\partial {\bf m}}{\partial x_{\mu}} 
\right)^2 
\label{action_rescaled} 
\end{eqnarray}
with coupling constant $g_0$ = 
$\sqrt{2}\Lambda aS^{-1}(J_1+J'_1)^{1/2}
(1/J_0 + 1/J'_0)^{1/2}$. 
      By RG equations up to one-loop approximation, 
the quantum phase transition from the 
AF ordered (N\'eel) state to a disordered state 
takes place at $g_0 = 4\pi$. 
      Rewriting this, the phase boundary 
in the space of the exchange parameters is given by 
\begin{eqnarray}
(J_1+J'_1) \left( \frac{1}{J_0}+\frac{1}{J'_0} \right) 
= \frac{2}{\lambda} \ \ {\rm with} \ \ \lambda  
\equiv \left( \frac{\Lambda a}{2\pi S} \right)^2 . 
\label{critical_general} 
\end{eqnarray}
      Parameter $\lambda$ represents the strength of 
quantum effect; $\lambda = 0$ 
in the classical spin limit. 

      To make the NLSM method complete, we determine 
the cutoff $\Lambda$ by considering the number 
of degrees of freedom for the square lattice. 
      The variable ${\bf m}$ is originally defined 
for each block of size $2a \times 2a$  
(Fig.~\ref{plaquette}(b) and Eq.~(\ref{transform})), 
and is taken a continuum limit. 
      Hence the correspondence of the momentum spaces 
is expressed as $(\pi/a)^2$ = $\pi \Lambda^2$, 
or the cutoff is given by 
$\Lambda = \sqrt{\pi}/a$. 
      Thus Eq.~(\ref{critical_general}) unambiguously 
determines the phase boundary between 
the ordered and the disordered phases. 

      In the uniform limit 
($J_1$ = $J'_1$, $J_2$ = $J'_2$ = $J''_2$), 
the system depends only on 
{\it frustration parameter} 
$\alpha \equiv$ $J_2/J_1$ and Eq.~(\ref{critical_general}) 
is reduced to $\alpha$ = $\frac{1}{2}-\lambda$. 
      Hence, for $S=\frac{1}{2}$ with 
$\Lambda = \sqrt{\pi}/a$, 
the critical value for $\alpha$ is given as 
$\alpha_c \simeq 0.18$. 
      Thus the NLSM method succeeds in producing 
a critical value satisfying 
$0 < \alpha_c < \frac{1}{2}$ 
without any additional assumption or interpretation. 
      The value is smaller than $\sim$0.4 estimated by 
the QMC simulation \cite{Sorella,Capriotti1}. 
      The deviation reflects the difference 
between the dispersions for spin-wave excitations 
in the lattice and the continuum models, 
and may be reduced by adjusting the cutoff. 
      Since we aim at inspecting the continuity of a phase, 
we do not need such a phenomenological adjustment. 

      In the limit of no frustration 
($J_2$ = $J'_2$ = $J''_2$ = 0), the plaquette distortion 
may cause an order-disorder transition. 
      We denote the strength of the distortion by 
{\it distortion parameter} $\gamma$ defined 
as $J'_1$ = $(1-\gamma) J_1$. 
      Then Eq.~(\ref{critical_general}) produces 
the critical value  
$\gamma_c = 2 - \lambda^{-1} + 
\sqrt{\lambda^{-2} - 2\lambda^{-1}}$. 
      This value decreases from 1 to 0 as $\lambda$ 
increases from 0 to $\frac{1}{2}$. 

\begin{figure}[b] 
\begin{center}\leavevmode
\includegraphics[width=0.9\linewidth]{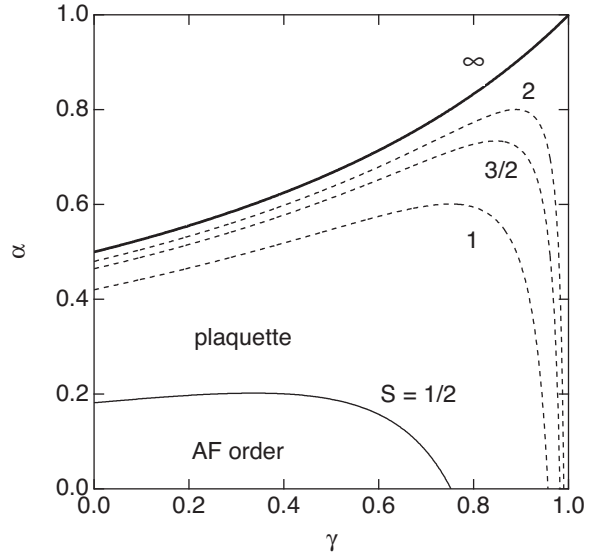}
\end{center}
\vskip -0.4cm
\caption{Phase diagram 
in the space of distortion parameter $\gamma$ 
and frustration parameter $\alpha$ (=$J_2/J_1$). 
      The bold solid line for $S$=$\infty$ separates 
the classical N\'eel and the classical colinear phases. 
      The region between the bold solid and the thin solid 
lines is the gapful plaquette phase for $S$=$\frac{1}{2}$. 
      The phase boundaries for $S$ = 1, $\frac{3}{2}$ 
and 2 are also shown by dashed lines. 
} 
\label{phase}
\end{figure}

      We now examine the continuity of the ground state 
between both the limits above. 
      To be concrete, we parameterize the exchange 
constants as 
$J'_1$ = $(1-\gamma) J_1$, 
$J'_2$ = $(1-\gamma)^{2} J_2$ and 
$J''_2$ = $(1-\gamma) J_2$ for 
$0 \le \gamma < 1$. 
      Equation (\ref{critical_general}) for the phase 
boundary is reduced to a simple form as 
$\alpha = (2-\gamma)^{-1}$ $-$ 
$\frac{1}{2}\lambda (2-\gamma) (1-\gamma)^{-1}$. 
      The phase diagram in the $\gamma$-$\alpha$ 
parameter space is shown in Fig.~\ref{phase}. 
      The bold line with $S$=$\infty$ is 
the classical  phase boundary between 
the N\'eel and the colinear phases~\cite{classical}. 
      The phase boundary of $S$=$\frac{1}{2}$ 
between the gapful and the gapless phases 
for variable {\bf m} is the thin solid line; 
      the state above is gapful, while 
that below is gapless corresponding to 
the N\'eel (AF) ordered state. 
      Boundaries for other spin magnitudes $S$ 
are also shown by dashed lines. 

      The gapful region of {\bf m} in Fig.~\ref{phase} 
extends continuously from the uniform limit 
on the $\alpha$-axis ($\gamma$=0) to 
the limit of no frustration 
on the $\gamma$-axis ($\alpha$=0). 
      Remembering that fields 
${\bf L}_0$, ${\bf L}_1$ and ${\bf L}_2$ are gapful, 
there is no gapless excitation throughout the region 
whether it is triplet or singlet. 
      Hence, the whole gapful region in Fig.~\ref{phase} 
is a single disordered phase. 
      In particular, the phase continues to the point 
of ($\gamma$, $\alpha$) = (1, 0) \cite{isolated}. 
      Hence {\it a disordered ground state on the 
$\alpha$-axis finally continues to the ground state 
of the assembly of isolated plaquettes.} 

      Thus there remain two possibilities for a disordered 
ground state of the uniform $J_1$-$J_2$ model, which 
is on the $\alpha$-axis in the phase diagram 
(Fig.~\ref{phase}). 
      First, the translational symmetry 
may be spontaneously broken; 
then the ground states are fourfold degenerate and 
one of them continues to the ground state 
at  ($\gamma$, $\alpha$) = (1, 0). 
      Second, the symmetry may not be spontaneously broken; 
then the ground state is unique and is a uniform RVB state 
with strong fluctuations of plaquette-singlets. 
      However, the second possibility is excluded by 
Oshikawa's commensurability condition~\cite{Oshikawa}. 
       Applying it to the present case, 
a uniform ground state with triplet excitation gap 
must be accompanied with  
other gapless excitations like singlet ones. 
      Such gapless excitations do not exist as we have 
already shown. 
      We therefore conclude that 
{\it the disordered ground states for the uniform 
$J_1$-$J_2$ model are fourfold degenerate 
plaquette states with spontaneously broken
translational invariance.} 

      Finally, we discuss possible experiments to 
detect the plaquette state for materials with 
$J_2/J_1 \sim 0.45$ which will be hopefully 
found in future. 
      In a realistic layered material, the uniform 
$J_1$-$J_2$ model is accompanied with at least 
weak three dimensionality. 
      Hence, at a finite temperature, the system will 
spontaneously break the translational symmetry 
to fall into a plaquette phase. 
      The appearance of a spin-gap at the temperature 
will be observed. 
      The characteristics of the plaquette state 
appear in the dispersion relation, which reflects 
the invariance for the translations of $2a$ 
in the $x$- and the $y$-directions. 
      They will be observed in neutron scattering 
experiments. 
      In a material where the spin system weakly 
interacts with the lattice, 
the spontaneous symmetry breaking 
for the spin degrees of freedom may 
induce a plaquette-type lattice distortion, 
which will be observed by X-ray diffraction. 
      Such a distorted system may correspond to a point 
deviated from the $\alpha$-axis in the plaquette phase 
of Fig.~\ref{phase}. 

      This work is partially supported by the Grant-in-Aid
for  Scientific Research from the Ministry of Education,
Culture, Sports, Science and Technology of Japan.  



\begin{references}
\bibitem{classical} 
      The present analysis is performed 
in the classical N\'eel region, $J_0+J'_0>0$.
\bibitem{isolated} 
      The NLSM method is not applicable to this point, 
since the continuum approximation is unsuitable 
to the isolated plaquettes. 
      However the ground state is not singular 
at the point. 
      In fact, the ground state is the direct product of the 
unique singlet ground states for the isolated plaquettes, 
and the excitation gap is of order $J_1$. 
      As $\gamma$ decreases from 1 on the $\gamma$-axis, 
the ground state remains nondegenerate with 
the plaquette symmetry at least for $J'_1 \ll J_1$.
\end{references}
\end{document}